\documentclass[aps,epsfig,prb,twocolumn]{revtex4}
\usepackage{amsmath}
\usepackage{graphicx}

\def\nn{\nonumber}

\begin{document}

\title{Effects of Screening on Propagation of Graphene Surface Plasmons}

\author{Ken-ichi Sasaki}
\email{sasaki.kenichi@lab.ntt.co.jp}
\affiliation{NTT Basic Research Laboratories, NTT Corporation,
3-1 Morinosato Wakamiya, Atsugi, Kanagawa 243-0198, Japan}

\author{Norio Kumada}
\affiliation{NTT Basic Research Laboratories, NTT Corporation,
3-1 Morinosato Wakamiya, Atsugi, Kanagawa 243-0198, Japan}

\date{\today}

\begin{abstract}
 Electromagnetic fields bound tightly to charge carriers in a
 two-dimensional sheet, namely surface plasmons, are shielded by
 metallic plates that are a part of a device.
 It is shown that for epitaxial graphenes,
 the propagation velocity of surface plasmons is suppressed
 significantly through a partial screening of the electron charge 
 by the interface states.
 On the basis of analytical calculations of the electron lifetime
 determined by the screened Coulomb interaction, 
 we show that the screening effect gives results in
 agreement with those of a recent experiment.
\end{abstract}

\maketitle

Plasmons, which consist of carriers and electromagnetic fields, 
are the principal elements of excited states in solids.~\cite{Mahan2000}
When carriers are confined in a two-dimensional layer,
surface plasmons can exist. 
The electromagnetic fields appear outside the layer and can be
sensitive to the screening effect provided, for example, by a metallic
plate that is a part of a device,~\cite{Ando1982}
which is not so obvious for other excited states in solids, such as
electrons and phonons.
A device composed of a two-dimensional sheet of carbons,
graphene,~\cite{novoselov05,zhang05} provides a great opportunity to
study this sensitivity of surface plasmons, as was demonstrated by
a recent time-resolved experiment, which we review below.

\begin{figure*}[htbp]
 \begin{center}
  \includegraphics[scale=0.6]{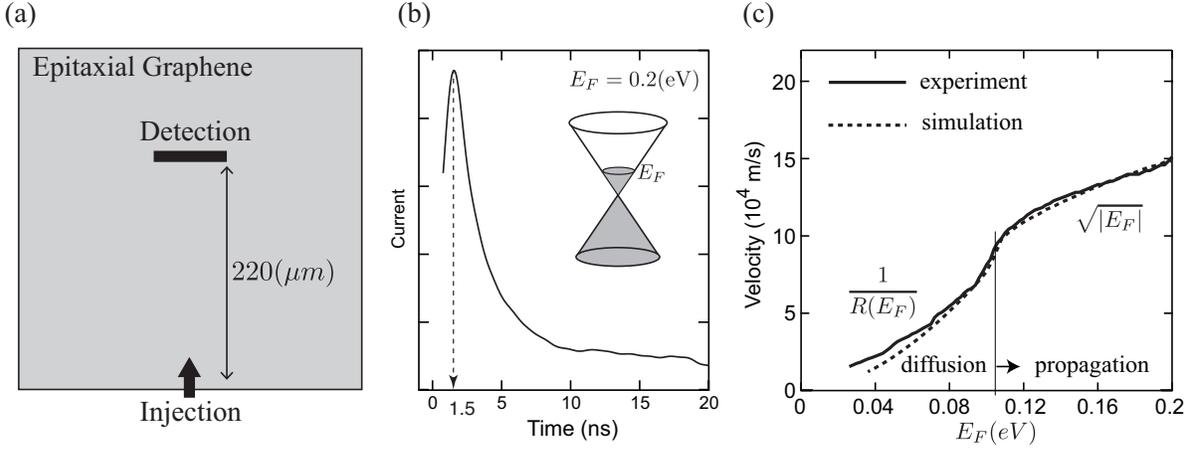}
 \end{center}
 \caption{(a) Schematic of a time-resolved transport experiment on
 epitaxial graphene.~\cite{Kumada2014}
 Because the experiments were performed at 1.5 K, the finite temperature
 effect~\cite{Klimchitskaya2014} can be safely ignored.
 (b) The waveform of the current at the detection gate is given as a
 function of time. The details of the waveform are dependent on $E_F$,
 which is controlled by a metal top gate covering the entire sample. 
 This plot corresponds to $E_F\simeq 0.2$ eV. 
 (c) The $E_F$ dependence of the propagation velocity.}
 \label{fig:exp_data}
\end{figure*}

Figure~\ref{fig:exp_data}(a) is the schematic of a transport experiment
performed by Kumada {\it et al} on graphene grown by SiC sublimation.~\cite{Kumada2014}
After applying a current pulse with a frequency of a few GHz at the
injection gate on epitaxial graphene,
they observed the current induced
at the detection gate located approximately 220 $\mu m$ from the injection gate.
Figure~\ref{fig:exp_data}(b) shows an example of the current observed as
a function of time.
The waveform has a peak structure at 1.5 ns, which enabled the authors to define 
the propagation velocity of a pulse as the propagation distance divided by the peak
time, i.e., $220 \mu m/1.5 {\rm ns}\simeq 15\times 10^4$ m/s.
The details of a waveform, such as peak time, depend on the Fermi energy
position $E_F$, which was controlled using a metal top gate in their
experiment.
As a result, they were able to find the $E_F$ dependence of the velocity shown
by the solid curve in Fig.~\ref{fig:exp_data}(c).
The velocity decreases as the Fermi energy approaches the Dirac point
$E_F=0$ eV.
For a wide range of $E_F$ the velocity is one order of
magnitude smaller than the electron Fermi velocity $v_F \simeq 10^6$ m/s.
Such a slow charge propagation in a gated graphene on SiC has been
observed also for edge magnetoplasmons.~\cite{Kumada2013}
The velocity in a device without a top gate was observed to be one or
two order of magnitude larger than $v_F$, suggesting that the
presence/absence of the gate strongly affects the plasmon transport.

In this paper we provide a theoretical basis that is useful for
studying the propagation velocity of surface plasmons in graphene, 
while paying particular attention to the effect of a metal gate on the
transport properties.~\footnote{
We assume that in this paper, the dielectric constant of a metal top
gate is $-\infty$ (i.e., a perfect electric conductor for modeling a
metal top gate), which is valid at GHz frequencies. 
The validity of this assumption needs to be checked for frequencies
higher than tens of terahertz.} 
We will show that in the absence of a metal gate,
plasmons propagate faster than the electrons.
In the presence of a metal gate,
the propagation velocity is much slower than $v_F$ when 
the screening effect provided by interface states is taken into account.
Furthermore, slow-moving surface plasmons
undergo a strong diffusion when $E_F$ is near the Dirac point, 
which explains the drop at $E_F \simeq 0.1$ eV seen in
Fig.~\ref{fig:exp_data}(c).

\begin{figure}[htbp]
 \begin{center}
  \includegraphics[scale=0.5]{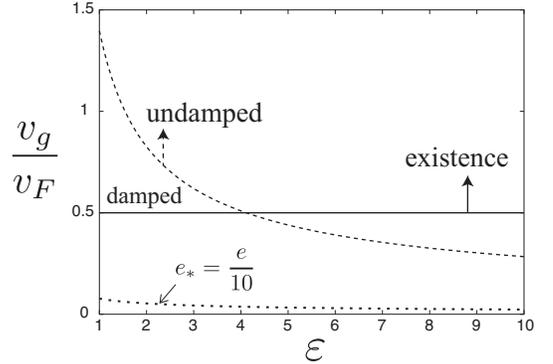}
 \end{center}
 \caption{The lower limit of the group velocity of the plasmons in
 graphene without a nearby metal top gate.
 The plasmons cannot exist (are damped) when $v_{g}/v_F$ is
 located below the solid line (dashed curve).
 Screening lowers the dashed curve:
 the dotted curve is when $e_{*}=e/10$.
 }
 \label{fig:velocity}
\end{figure}

We begin by showing that
the group velocity of plasmons in graphene without a metal gate cannot
be lower than $v_F/2$.
The plasmon dispersion is derived from the zero value of the real part
of the dielectric constant 
\begin{align}
 \varepsilon_{E_F}(q,\omega) =
 1 - v_q {\rm Re} \Pi_{E_F} (q,\omega) = 0,
 \label{app:1}
\end{align}
where $v_q$ is the Coulomb potential.~\cite{Mahan2000,Wunsch2006,hwang07}
In the absence of a metal top gate,
$v_q = 2\pi e^2/\varepsilon q$, where $\varepsilon$ is the permittivity of a
surrounding medium, $q$ is the wavevector magnitude, and $e$ is
electron charge magnitude in vacuum ($e^2=1.44$ eV$\cdot$nm).
$\Pi_{E_F} (q,\omega)$ is the polarization function, which is a function
of $q$, frequency $\omega$, and $E_F$.
Although the polarization function for doped graphene has been
calculated in several
papers,~\cite{Wunsch2006,hwang07,sasaki12_migration} we show it in 
Appendix~\ref{app:A} for clarity.
Since $v_q > 0$, the solution of Eq.~(\ref{app:1}) exists only when 
${\rm Re} \Pi_{E_F} (q,\omega)>0$ is satisfied.
It can be shown that 
${\rm Re} \Pi_{E_F} (q,\omega)>0$ when $\omega>v_Fq$
and ${\rm Re} \Pi_{E_F} (q,\omega)<0$ when $\omega<v_Fq$, so that
plasmons exist only when $\omega>v_Fq$.~\footnote{
This behavior of ${\rm Re} \Pi_{E_F}(q,\omega)<0$ when $\omega<v_Fq$
originates from the fact that softening dominates hardening.
Softening/hardening here refers to the negative/positive contributions
to the real part of the polarization function. 
The significance of each contribution depends on the matrix element for
the interaction being considered.~\cite{sasaki12_migration} 
With the Coulomb interaction, 
the matrix element is at its maximum (minimum) value for forward
(backward) scattering (as shown by 
$1+ \cos\left(\Theta_{\bf k'}-\Theta_{\bf k}\right)$), by which the
contribution of the forward (backward) scattering that causes softening
(hardening) is enhanced.
As a result, softening dominates hardening 
so that ${\rm Re} \Pi_{E_F}(q,\omega)<0$ when $\omega<v_Fq$.
}
In the literature,
$\omega<v_Fq$ is referred to as an electron-hole continuum 
or an intraband single-particle excitation (or SPE$_{\rm intra}$) region, 
where plasmons do not exist.
When $\omega>v_Fq$, ${\rm Re} \Pi_{E_F} (q,\omega)$ is approximated in
the $q\to 0$ limit by
\begin{align}
 {\rm Re} \Pi_{E_F} (q,\omega) \simeq \frac{|E_F|}{\pi}
 \left(\frac{q}{\hbar \omega}\right)^2.
 \label{app:2}
\end{align}
On combining Eq.~(\ref{app:1}) with Eq.~(\ref{app:2}),
we obtain the plasmon frequency~\cite{Wunsch2006,hwang07}
\begin{align}
 \omega_{pl}(q,E_F) = \frac{1}{\hbar}
 \sqrt{\frac{2e^2 q |E_F|}{\varepsilon}}.
 \label{eq:dis_0}
\end{align}
The $q$ dependence of $\omega_{pl}$, namely $\sqrt{q}$, is common to
two-dimensional electron gas (2DEG) systems.~\cite{stern67}
The existence of plasmons requires that the frequency satisfies 
\begin{align}
 \omega_{pl}(q,E_F) > v_F q.
 \label{eq:exist}
\end{align}
Putting Eq.~(\ref{eq:dis_0}) into this condition,
we have
\begin{align}
 \frac{1}{\hbar}
 \sqrt{\frac{2e^2 |E_F|}{\varepsilon q}} > v_F.
 \label{eq:cond-2}
\end{align}
Because the group velocity is defined by
\begin{align}
 v_{g}(q,E_F) \equiv \frac{\partial \omega_{pl}(q,E_F)}{\partial q}
 =\frac{1}{2\hbar}
 \sqrt{\frac{2e^2|E_F|}{\varepsilon q}},
 \label{eq:pl-vel}
\end{align}
it is shown that by combining Eq.~(\ref{eq:cond-2}) with Eq.~(\ref{eq:pl-vel})
the plasmon group velocity has the lower limit
\begin{align}
 v_{g}(q,E_F) > \frac{v_F}{2}.
 \label{eq:vel-bound-1}
\end{align}
This lower limit of the group velocity
does not depend on $\varepsilon$, $q$, $E_F$, or
$e^2$, whereas the factor $1/2$ reflects the exponent of 
$q$ in the dispersion relation. 
The solid line in Fig.~\ref{fig:velocity} shows the lower limit.
The actual group velocity must be located above the solid line, as indicated
by the vertical arrow. 
It is also straightforward to show that 
the group velocity of an undamped plasmon will be located above the
dashed curve (see Appendix~\ref{app:B} for details). 

The conditions for the existence of plasmons 
and for plasmons to be undamped
give the lower limit of the propagation velocity,
while there is no condition that specifies the upper limit.
This result suggests that the propagation velocity of the plasmons is
generally high.
For example, it is shown by eliminating $q$ from Eq.~(\ref{eq:pl-vel})
using Eq.~(\ref{eq:dis_0}) that 
\begin{align}
 v_g(\omega_{pl}, E_F) = \frac{e^2}{\hbar \varepsilon} \frac{|E_F|}{\hbar \omega_{pl}}.
 \label{eq:vp_sub}
\end{align}
When $\omega_{pl}=10$ GHz, $E_F=0.1$ eV, and $\varepsilon =10$,
we have $v_{g} \simeq 6 \times 10^7$ m/s.

\begin{figure}[htbp]
 \begin{center}
  \includegraphics[scale=0.45]{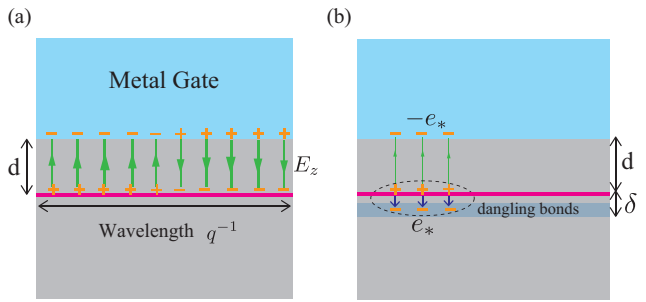}
 \end{center}
 \caption{(color)
 Cross-section of a device consisting of a graphene sheet (red) and a
 metal gate (blue). The gray regions represent a dielectric (dielectric
 constant $\varepsilon$).
 (a)
 The arrows represent the electric fields ${\bf E}=(0,0,E_z)$
 for non-radiative (acoustic) plasmons satisfying $q^{-1} \gg d$.
 We omit to draw an exponential decaying electric field which appears
 underneath the graphene sheet for clarity.
 (b) Interface states near to graphene sheet, including the dangling
 bond states ($\delta \ll d$), efficiently screen the electron charge.
 }
 \label{fig:gim}
\end{figure}

When a metal plate is placed at a distance, $d$, from
a graphene sheet as shown in Fig.~\ref{fig:gim}(a), 
we have a metal-insulator-graphene device. 
Nakayama showed that surface plasmons exist for such a
device.~\cite{nakayama74}
The dispersion relation is given by 
\begin{align}
 \omega_{s}(q) = \sqrt{\frac{2\pi \sigma_0 \tau^{-1}}{\varepsilon}}
 \sqrt{\frac{q}{1+\coth(qd)}},
 \label{eq:freq_qd}
\end{align}
where $\sigma_0$ is the static conductivity and $\tau$ is the relaxation
time.~\footnote{
It is assumed that in deriving Eq.~(\ref{eq:freq_qd}) 
the dynamical conductivity $\sigma_\omega$ is approximated by 
$i\sigma_0/\omega \tau$, which is a direct consequence of the Drude
model, $\sigma_\omega=\frac{\sigma_0}{1-i\omega \tau}$,
with the condition $\omega \tau \gg 1$. 
Note that when the imaginary part of the dynamical conductivity is
positive as shown above, only a transverse magnetic (TM) mode can exist.
Meanwhile when the imaginary part of the dynamical conductivity is
negative or in the presence of an external magnetic field, 
a transverse electric (TE) can appear.~\cite{nakayama74}
Mikhailov and Ziegler point out that the imaginary part of the dynamical
conductivity of graphene can be negative for a special
frequency,~\cite{Mikhailov2007}
because an interband transition contributes to
the dynamical conductivity, while the Drude model only accounts for an
intraband transition.
As a result, they predict that graphene can support a TE mode 
for a special frequency (even without an external magnetic field).
Another TE mode propagating at the speed of light 
is reported by Bordag and Pirozhenko,~\cite{Bordag2014}
but this can exist only when $E_F=0$.
}
In the absence of a metal gate (when $d \to \infty$), we can reproduce
Eq.~(\ref{eq:dis_0}) from Eq.~(\ref{eq:freq_qd}) using 
the Einstein relation,~\cite{Ando2008}
\begin{align}
 \sigma_0=e^2 v_F^2 \tau D(E_F),
 \label{eq:Einstein}
\end{align}
where $D(E_F)=2|E_F|/\pi(\hbar v_F)^2$ is the density
of states of graphene.
Thus, Eq.~(\ref{eq:freq_qd}) is a general result that includes
Eq.~(\ref{eq:dis_0}) as the limiting case.
In the presence of a metal gate,
the $\sqrt{q}$-dependence is lost for long-wavelength modes
(or when $q^{-1}\gg d$) and $\omega_{s}$ exhibits a linear dependence on
$q$ as 
\begin{align}
 \omega_{s}(q) = 
 \sqrt{\frac{2\pi \sigma_0 \tau^{-1}d}{\varepsilon}} q, \ \ (qd
 \ll 1).
 \label{eq:freq_qd_limit}
\end{align}
Then the group velocity is given by 
\begin{align}
 v \equiv \frac{\partial \omega_{s}(q)}{\partial q}
 =
 \sqrt{\frac{2\pi \sigma_0 \tau^{-1}d}{\varepsilon}}.
 \label{eq:vel_low}
\end{align}
The electric fields of surface plasmons have their principal component
normal to the graphene sheet ${\bf E}=(0,0,E_z)$, as shown in
Fig.~\ref{fig:gim}(a).
This field configuration is obtained by solving Maxwell's equations for
electromagnetic fields (see Ref.~\onlinecite{nakayama74} for details).
The field configuration is in sharp contrast to that in the absence of a
top gate (when $d \to \infty$), where the electric fields have components both normal and
parallel to the graphene sheet as ${\bf E}=(E_x,0,E_z)$ where
$E_x(x,z,t)=E e^{i(kx-\omega t)-\alpha |z|}$
and $E_z(x,z,t)=ikE_x(x,z;t)/\alpha$ with 
$\alpha \equiv \sqrt{k^2-\varepsilon\omega^2/c^2}$ 
(see Ref.~\onlinecite{nakayama74} for details).
Because $d$ is 200 nm and the condition $qd\ll 1$ is satisfied in
Ref.~\onlinecite{Kumada2014}, the excitation described by 
Eq.~(\ref{eq:vel_low}) is considered to be that observed in the
experiment in the presence of a metal top gate.
However, the application of the Einstein relation
Eq.~(\ref{eq:Einstein}) to Eq.~(\ref{eq:vel_low}) gives 
\begin{align}
 v = \frac{e}{\hbar} \sqrt{\frac{4|E_F|d}{\varepsilon}}.
 \label{eq:velsw}
\end{align}
The velocity predicted from Eq.~(\ref{eq:velsw}) with $\varepsilon=4$
and $d=200$ nm is $v=25\sqrt{|E_F|/{\rm eV}}\times 10^6$ m/s, which is
two orders of magnitude larger than that observed in the
experiment (see Fig.~\ref{fig:exp_data}(c)).

The discrepancy between the predicted and experimental values of
velocity can be accounted for by a modification of the Einstein relation
caused by a strong (but not perfect) screening effect
produced by interface (trap) states.
In an epitaxial graphene device grown on SiC, the interface states are
naturally realized by the dangling bond states at the SiC substrate 
[see Fig.~\ref{fig:gim}(b)].~\cite{Zebrev2011,takase12}
When the (positive) charge $e$ exists in the graphene sheet, 
a screening charge with approximately $-e$ is induced on the dangling
bond states.
Meanwhile, the screening effect of the interface states is not
perfect, and a (positive) charge with magnitude $e_*$ remains in the
capacitor consisting of the graphene sheet and dangling bond states, as
shown schematically in Fig.~\ref{fig:gim}(b).~\cite{Luryi1988}
This charge induces (negative) screening charge with $-e_*$ on the metal
top gate.
If surface plasmons consist of particles with charge magnitude $e_{*}$,
we can expect Eq.~(\ref{eq:Einstein}) to be defined by replacing $e$ 
with the screened charge $e_*$ as 
\begin{align}
 \sigma_0 = e_{*}^2 v_{\rm F}^2 \tau D(E_F).
 \label{eq:Einstein_new}
\end{align}
The corrected velocity is given by the application of
Eq.~(\ref{eq:Einstein_new}) to Eq.~(\ref{eq:vel_low}) as
\begin{align}
 v = \frac{e_{*}}{\hbar} \sqrt{\frac{4|E_F|d}{\varepsilon}}.
 \label{eq:velsw_screen}
\end{align}

The value of $e_{*}$ can be roughly estimated by an extension of the
result of Luryi,~\cite{Luryi1988} in which $e_*$ is expressed in terms
of the quantum capacitance of the interface states $C_i$ [$=e^2
\gamma$] and geometrical capacitance $C_d$ [$=\varepsilon/d$] as 
\begin{align}
 e_{*} \simeq \frac{C_d}{C_i+C_d}e,
\end{align}
in the static limit.
When we adopt the value $\gamma=0.37$
eV$^{-1}\cdot$nm$^{-2}$ obtained by Takase {\it et al.},~\cite{takase12}
$e_{*}/e\simeq 0.036$ for $d=200$ nm and $\varepsilon =4$.
This value is in agreement with the experiment.~\footnote{Although
Eq.~(\ref{eq:freq_qd}) is obtained by solving Maxwell's equations for
electromagnetic fields in the framework of classical mechanics,~\cite{nakayama74}
when we consider it in quantum mechanics, we can conclude that 
the frequency does not obey the plasmon existence condition
Eq.~(\ref{eq:exist}) when $v<v_F$ holds as a result of screening.
Indeed, an analysis based on the polarization function suggests that when
$v<v_F$, the mean lifetime of the plasmon is of the order of
a femtosecond (see Appendix~\ref{app:C} for details), and the plasmons quickly
decay into intraband single-particle electron-hole pairs. 
In this case, we interpret the plasma surface waves as a density
fluctuation consisting of single particle electron-hole
pairs.~\cite{Allen1997}
Then it is reasonable to consider that the peak time $t^*$ in the
waveform is limited by the quasi-particle lifetime 
(see Appendix~\ref{app:D} for details).
}
The advantage of incorporating screening is that 
as long as the interface states near to graphene are taken into
account through the modification of the Coulomb potential in
Eq.~(\ref{app:1}) as 
\begin{align}
 v_q=\frac{2\pi e_{*}^2}{q},
\end{align}
the conclusion obtained in the absence of a metal top gate is valid even
in the presence of the interface states since the lower limit stems
from the $\sqrt{q}$-dependence of $\omega_p$ and is independent of the
electron charge as shown in Eq.~(\ref{eq:vel-bound-1}). 
This result is also consistent with the experiment.


Propagation velocity can be suppressed
by resistivity $R$, which is not taken into account in
Eq.~(\ref{eq:velsw_screen}).
To investigate the effect of $R$ on the propagation velocity,
we can adopt an $RLC$ circuit model 
introduced by Burke {\it et al.} for studying plasmons in a 2DEG
system.~\cite{burke00}
The use of this model was motivated by the fact that 
the electric fields in the dielectric shown in
Fig.~\ref{fig:gim}(b) are similar to those in a
waveguide, for which the wave propagation is described by an $RLC$
circuit model.
In this model, $C$ and $L$ correspond to $C_d$ and the kinetic
inductance of graphene, respectively. 
In Ref.~\onlinecite{Kumada2014} 
we simulated the time evolution of the pulse using the Runge-Kutta
method and obtained the waveform at the detector.
By following the procedure used in the experiment, 
we determined the propagation velocity of the pulse in terms of the peak
time and obtained the dashed curve in Fig.~\ref{fig:exp_data}(c). 
Our simulation reproduces the experimental result satisfactorily.
The effect of $R$ on the propagation velocity can be examined
analytically in terms of the continuum approximation of the
$RLC$ circuit model given by the telegrapher's
equation,~\cite{Kumada2014,doetsch71,sonnenschein00}
\begin{align}
 \left[ \frac{\partial^2}{\partial t^2} - v^2 \nabla^2 +\frac{R}{L}
 \frac{\partial}{\partial t} \right] E_z({\bf r},t) = 0,
 \label{eq:TE}
\end{align}
where
\begin{align}
 v = \frac{1}{\sqrt{LC_d}},
 \label{eq:LCv}
\end{align}
and inductance $L$ is given from Eqs.~(\ref{eq:vel_low}) and
(\ref{eq:Einstein_new}) by 
\begin{align}
 L = \frac{\tau}{2\pi \sigma_0} =\frac{1}{2\pi e_{*}^2v_F^2 D(E_F)}.
 \label{eq:LC}
\end{align}
The solution may be constructed from the Green's
function of the Klein-Gordon equation,
\begin{align}
 \left[\frac{\partial^2}{\partial t^2} -v^2 \nabla^2 + m^2 \right] \phi({\bf r},t)=0,
\end{align}
with a negative mass squared $m^2=-q_c^2 v^2$ where $q_c$ is the damping
factor, 
\begin{align}
 q_c \equiv \frac{R}{2} \sqrt{\frac{C_d}{L}},
 \label{eq:qc}
\end{align}
because the telegrapher's equation is reproduced
from the Klein-Gordon equation by setting $\phi({\bf r},t)= e^{q_c
vt}E_z({\bf r},t)$.
The retarded Green's function of the Klein-Gordon equation is well-known
and written as
$\Delta_R({\bf r},t)= \theta(t) \Delta({\bf r},t)$ where
\begin{align}
 & \Delta({\bf r},t)=\frac{{\rm sgn}(t)}{2\pi}\times \nn \\
 & \left[\delta(t^2-\frac{|{\bf r}|^2}{v^2})-\frac{m}{2}\theta(t^2-\frac{|{\bf r}|^2}{v^2})
 \frac{J_1(m\sqrt{t^2-\frac{|{\bf r}|^2}{v^2}})}{\sqrt{t^2-\frac{|{\bf
 r}|^2}{v^2}}} \right],
\end{align}
and $J_1(x)$ is the Bessel function of the first kind.
Thus, by specifying the initial condition
$E_z(0,t)$, the solution of the telegrapher's equation 
is written as $E_z(x,t) = E_p(x,t) +E_d(x,t)$ for $t > x/v$ 
where~\cite{doetsch71,sonnenschein00}
\begin{align}
 &E_p(x,t)=  e^{-q_c x} E_z(0,t-\frac{x}{v}),
 \label{eq:Ep} \\
 &E_d(x,t)= q_c x \int_{\frac{x}{v}}^t e^{-q_c v t'}
 \frac{I_1(q_c v
 \sqrt{t'^2-\frac{x^2}{v^2}})}{\sqrt{t'^2-\frac{x^2}{v^2}}} E_z(0,t-t')
 dt'.
 \label{eq:Ed}
\end{align}
Here, we used $I_1(x)=-iJ_1(ix)$ where $I_1(z)$ is the modified Bessel
function of the first kind.
Here, $E_p(x,t)$ is an exponentially decaying signal 
that propagates at a speed $v$, and 
$E_d(x,t)$ expresses diffusion.

\begin{figure}[htbp]
 \begin{center}
  \includegraphics[scale=0.35]{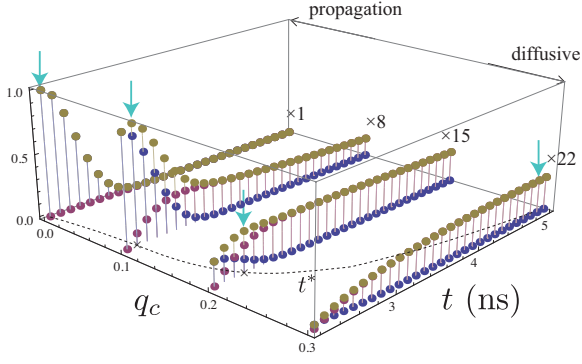}
 \end{center}
 \caption{(color) The $q_c$-dependence of the waveform at $x=220$ $\mu m$.
 In this plot we assume $T=400$ ps and $v=10^5$ m/s, and $q_c$ is given
 in the units of $(10 \mu m)^{-1}$.
 The peak time of the $E_p$ component (blue) is $t=2.2$ ns, while that of the $E_d$
 component (red) increases with increasing $q_c$.
 The sum of the two components, $E_z$, is referred by yellow.
 Arrows represent peak time.
 }
 \label{fig:waveform}
\end{figure}

The effect of $R$ on the plasmon propagation 
is most clearly visualized at the drop in the peak velocity observed
below $E_F\simeq 0.1$ eV in Fig.~\ref{fig:exp_data}(c) 
which is due to the dominance of diffusion.
For the $\delta$-function initial pulse $E_z(0,t)=\delta (t)$,
it is shown that
by differentiating Eqs.~(\ref{eq:Ep}) and (\ref{eq:Ed}) with respect to $t$,
the time $t^*$ corresponding to the peak in the waveform 
is $t^*=x/v$ for $E_p(x,t)$ and $t^*\approx q_c x^2/3v$ for $E_d(x,t\gg
x/v)$.~\cite{sonnenschein00}
Thus, when $E_p$ dominates $E_d$ (propagation dominant), 
the peak velocity is given by $x/t^*=v$, on the other hand, 
when $E_d$ dominates $E_p$ (diffusion dominant) 
the peak velocity is suppressed
by the factor of $3/q_c x$ as $x/t^* \approx 3v/q_c x$. 
Hence, when diffusion dominates, the peak velocity exhibits the
$E_F$ dependence of $v/q_c$ ($\propto 1/R$), while 
when propagation dominates it exhibits the $v$ ($\propto \sqrt{|E_F|}$)
dependence.
Whether $E_d$ dominates $E_p$ can depend sensitively on the
value of $q_c$.
This should be examined for a more realistic initial pulse, namely 
for the Gaussian initial pulse $E_z(0,t)=\exp(-t^2/T^2)$ where $T=400$ ps.~\cite{Kumada2014}
We plot $E_p(x,t)$, $E_d(x,t)$, and $E_z(x,t)$ at $x=220$ $\mu m$ 
for different $q_c$ values in Fig.~\ref{fig:waveform}.
When $q_c =0$, the peak time is seen at $t^*=2.2$ ns, so the propagation
velocity $x/t^*$ is $10^5$ m/s, which is approximately equal to the
velocity at $E_F= 0.1$ eV in Fig.~\ref{fig:exp_data}(c).
In Fig.~\ref{fig:waveform}, it is seen that 
when $q_c <0.1$, $E_p$ dominates $E_d$, whereas
when $q_c >0.3$, $E_d$ dominates $E_p$.
The maximum amplitudes of $E_p$ and $E_d$ are similar 
when $q_c \simeq 0.2$. 
The peak time $t^*$ increases rapidly when $q_c$ changes very slightly
from 0.2 to 0.3. 
This means that the peak velocity decreases rapidly then,
which can explain that the velocity decreases rapidly below $E_F\simeq 0.1$
eV in Fig.~\ref{fig:exp_data}(c).
Indeed, when we adopt the values obtained in Ref.~\onlinecite{Kumada2014}: 
$R(E_F)=340+3.7\times 10^6/(22+(500E_F)^2)$ $\Omega$
and $\sqrt{C_d/L}=\sqrt{0.58|E_F|}\times 10^{-3}$ $\Omega^{-1}$,
$q_c$ changes from 0.218 to 0.283 when $E_F$ decreases a little from 0.1
to 0.08.

Since $E_d(x,t)$ is proportional to $q_c$, 
diffusion is suppressed by decreasing $q_c$, which may be realized
by decreasing $R$ or increasing $L$ [see Eq.~(\ref{eq:qc})].
Achieving a large $L$ (or small $R$) is also important in order to 
extend the relaxation time $\tau' \equiv 1/(vq_c)=2L/R$
or to suppress the damping caused by $\exp(-q_c v t)$ for $E_p(x,t)$.
However, it should be noted that since both $L$ and $R$ decrease as
$|E_F|$ increases, 
increasing $L$ by decreasing $|E_F|$ is incompatible with decreasing $R$.
On the other hand,
$L$ ($q_c$) is enhanced (suppressed) significantly by the screening
effect provided by the interface states.

To conclude, the effects of a metal top gate and interface states on the
plasmon transport have been revealed: 
the former provides linearly dispersed plasmons, while the latter renormalizes the effective charge.
In the absence of a metal top gate,
the propagation velocity of surface plasmons 
has a lower limit given by $v_F/2$.
This lower limit is a rigid consequence derived from the condition for
the existence of plasmons and independent of the electron charge in particular.
Thus, as long as the interface states are taken into
account as the origin of the partial screening effect 
(i.e., $v_q=2\pi e^2/q\to 2\pi e_*^2/q$), 
the conclusion is valid even in the presence of the interface states.
In the presence of a metal top gate, the lower limit may be ineffective due
to the modification of the dispersion relation of the surface plasmons
($\omega_{pg} \propto \sqrt{q} \to q$). 
For the linear dispersion, we could utilize the concept of inductance for
analyzing the velocity.
An analysis using the $RLC$ circuit model and telegrapher's equation
successfully explained the experimental results for the $E_F$ dependence
of the propagation velocity, which proves that the inductance is effectively
enhanced in the presence of a metal top gate.
We attributed the enhancement to the screening effect induced by the
interface states and found the idea to be consistent with the
electron lifetime.
A straightforward deduction from our results is that surface plasmons in
a device consisting of exfoliated graphene without interface states
experiences strong dumping and the propagation is severely suppressed.
In other words, epitaxial graphenes have an advantage over exfoliated
graphenes in realizing high inductance.

\section*{Acknowledgments}

We are grateful to Yasuhiro Tokura for helpful discussions.

\appendix

\section{Polarization function}\label{app:A}

In this appendix, we use $v$ for $v_F$ and $\mu$ for $E_F/\hbar$.
The polarization function is given by 
\begin{widetext}
\begin{align}
 & -{\rm Im} \Pi_\mu(q,\omega) = \nn \\
 & \frac{1}{2\pi} \frac{(vq)^2}{\sqrt{\omega^2-(vq)^2}} \theta_{\omega-v q} \left[
 \theta_{\frac{\omega-vq}{2}-\mu} \left\{F(1)-F(-1)\right\}
 + \theta_{\mu-\frac{\omega-vq}{2}} \theta_{\frac{\omega+vq}{2}-\mu}
 \left\{ 
 F(1)-F\left(\frac{2\mu-\omega}{vq}\right) \right\}
 \right] + \nn \\
 & \frac{1}{2\pi} 
 \frac{(vq)^2}{\sqrt{(vq)^2-\omega^2}} \theta_{vq-\omega}
 \left[ \theta_{\mu-\frac{vq-\omega}{2}}
 G\left( \frac{2\mu+\omega}{vq} \right)-
 \theta_{\mu-\frac{\omega+vq}{2}}
 G\left(\frac{2\mu-\omega}{vq}\right)
 \right]
 \label{eq:imPI}
\end{align}
\begin{align}
 & {\rm Re} \Pi_\mu(q,\omega) = -\frac{2\mu}{\pi} \nn \\
 &-\frac{1}{2\pi} 
 \frac{(vq)^2}{\sqrt{\omega^2 -(vq)^2}} 
 \theta_{\omega-vq}
\left[
 \theta_{\frac{\omega-vq}{2}-\mu}G\left(\frac{\omega-2\mu}{vq}\right)
+\theta_{\mu-\frac{\omega+vq}{2}} G\left(\frac{2\mu -\omega}{vq}\right) -G\left(\frac{\omega+2\mu}{vq}\right)
\right] \nn \\
 &-\frac{1}{2\pi} \frac{(vq)^2}{\sqrt{(vq)^2-\omega^2}} \theta_{vq-\omega}
 \left[
 \theta_{\frac{vq+\omega}{2}-\mu}\left\{ F(1)-F\left(\frac{2\mu-\omega}{vq} \right)\right\}
+ \theta_{\frac{vq-\omega}{2}-\mu} \left\{ F\left(1\right) - F\left(\frac{\omega+2\mu}{vq}\right)\right\}
\right]
\end{align}
\end{widetext}
where $\theta_x$ denotes the step function satisfying $\theta_{x\ge 0}=1$ and
$\theta_{x<0}=0$. 
The functions $F$ and $G$ are defined by 
\begin{align}
 & F(x)=\frac{1}{2} \left\{ x\sqrt{1-x^2} +
 \sin^{-1}(x) \right\}, \\
 & G(x)=
 \frac{1}{2} \left\{ x\sqrt{x^2-1} -\ln\left(x+\sqrt{x^2-1}\right) \right\},
\end{align}
respectively.
We showed a direct derivation of the above formula in Supplemental
Material of Ref.~\onlinecite{sasaki12_migration}, for which we need to
multiply $g_v/(2\pi \hbar v_F)^2$ with $g_v=2$.

\section{}\label{app:B}

The dashed curve in Fig.~\ref{fig:velocity} is based on the
inequality given by
\begin{align}
 v_{p}(q,E_F)>
 \frac{e}{2\hbar} 
 \sqrt{\frac{\hbar v_F + \frac{e^2}{2\varepsilon} + \sqrt{\hbar v_F
 \frac{e^2}{\varepsilon} +\left(\frac{e^2}{2\varepsilon}
 \right)^2}}{\varepsilon}}. 
 \label{eq:stacond}
\end{align}
This lower limit of the group velocity
depends on the values of $\varepsilon$ and $e$.
The dotted curve in Fig.~\ref{fig:velocity} is the plot when $e$ is
replaced with $e^*=e/10$.

Equation~(\ref{eq:stacond}) arises from the fact that 
plasmons can decay into the constituent (interband) electron-hole
pairs of the collective charge-density oscillations. 
The decay is suppressed (plasmons become undamped) when
\begin{align}
 |E_F| > \frac{\hbar \omega_{p}(q,E_F)+\hbar v_Fq}{2}
 \label{eq:cond-3}
\end{align}
holds, otherwise the decay of plasmons into single particle
electron-hole pairs is not negligibly small.~\cite{sasaki12_migration}
Mathematically, Eq.~(\ref{eq:cond-3})
is equivalent to a condition where the imaginary part of the
polarization function Eq.~(\ref{eq:imPI})
vanishes: ${\rm Im}\Pi_{E_F}(q,\omega_{p}(q))=0$
for $\omega_{p}(q)>v_Fq$.
The condition of Eq.~(\ref{eq:stacond}) can be obtained 
by putting Eq.~(\ref{eq:dis_0}) into Eq.~(\ref{eq:cond-3}) to obtain
\begin{align}
 |E_F| > \frac{q}{2} \left( \hbar v_F + \frac{e^2}{2\varepsilon} + \sqrt{\hbar v_F
 \frac{e^2}{\varepsilon} +\left(\frac{e^2}{2\varepsilon}
 \right)^2}\right),
\end{align}
and then by using Eq.~(\ref{eq:pl-vel}).

\section{}\label{app:C}

When $\omega \equiv v q < v_F q$ (or $v < v_F$) and 
Eq.~(\ref{eq:cond-3}) is satisfied,
the imaginary part of $\Pi_{E_F}(q,\omega)$ is written as 
\begin{align}
 & - {\rm Im} \Pi_{E_F}(q,\omega)= \frac{\hbar v_Fq}{4\pi \sqrt{1- \frac{\omega^2}{(v_Fq)^2}}}\nn \\
 & \times
 \left\{ 
 G\left(\frac{2|E_F|+\hbar \omega}{\hbar v_Fq}\right)
 - G\left(\frac{2|E_F|-\hbar \omega}{\hbar v_Fq}\right)
 \right\},
\end{align}
where $G(x)\equiv \left\{x\sqrt{x^2-1} - \log
\left(x+\sqrt{x^2-1}\right)\right\}/2$. Note that $dG(x)/dx=\sqrt{x^2-1}$.
According to the time-energy uncertainly relation, the mean
lifetime is approximated by
\begin{align}
 \tau \equiv -\frac{\hbar}{2 {\rm Im} \Pi_{E_F}(q,\omega)}
 \simeq \frac{v_F}{v} \sqrt{1-
 \left(\frac{v}{v_F}\right)^2} \frac{\pi \hbar}{|E_F|}.
\end{align}
The characteristic time scale of $\tau$ is of the order of a femtosecond
because $\tau\simeq 2.5\times 10^{-2}(v_{\rm
F}/v)|\varepsilon_{\rm F}|^{-1}$ fs ($\varepsilon_{\rm F}$ is in
units of eV), when $v \ll v_F$.

\section{}\label{app:D}

We examined the $d$ dependence of the electron's quasi-particle lifetime
determined by the Coulomb interaction to validate the assumption of screening.
The lifetime is given by the inverse of the imaginary part
of the electron selfenergy $\Sigma$ as $\tau_q= \hbar/2{\rm Im}\Sigma$.
We calculated ${\rm Im}\Sigma$ 
using the formula,~\cite{Quinn1958,Hawrylak1987,DasSarma2007}
\begin{align}
 & {\rm Im} \Sigma_{\bf k}(E_F>0) = 
 \int \frac{d^2{\bf k'}}{(2\pi)^2} 
 \left\{
 \theta(\xi_{\bf k}-\xi_{\bf k'})
 -\theta(\xi_{\bf k'}-E_F) \right\}\nn \\
 & \times \frac{1+ \cos\left(\Theta_{\bf k'}-\Theta_{\bf k}\right)}{2} 
 {\rm Im} \frac{v_{|{\bf k'}-{\bf k}|}}{\varepsilon_{E_F}(|{\bf k'}-{\bf k}|,\xi_{\bf k}-\xi_{\bf k'})},
\end{align}
where $\xi_{\bf k}=\hbar v_F |{\bf k}|$, $k_x - ik_y = |{\bf
k}|\exp(-i\Theta_{\bf k})$, and $v_q$ denotes the screened Coulomb
potential given by
\begin{align}
 v_q = \frac{4\pi e_{*}^2}{\varepsilon q(1+\coth(qd))}.
 \label{eq:new_vq}
\end{align}
Note that Eq.~(\ref{eq:velsw_screen}) may be obtained from
Eq.~(\ref{app:1}) with this $v_q$.~\cite{Principi2011}
A straightforward calculation shows that when 
$\xi_k\simeq E_F$, ${\rm Im} \Sigma_{\bf k}(E_F>0)$ is approximated by 
$\hbar v_F|\xi_k-E_F|/(16E_Fd)$.~\footnote{The linear dependence of
${\rm Im} \Sigma_{\bf k}$ on $\xi_k-E_F$ is in sharp contrast to the
result obtained in the absence of screening,~\cite{DasSarma2007} 
$-\frac{(\xi_k-E_F)^2}{16\pi E_F} \left\{
 \ln \left(\frac{(\xi_k-E_F)^2}{32E_F^2}\right) + 1 \right\}$.
}
As a result, we obtain
\begin{align}
 \tau_q = \frac{8E_Fd}{v_F|\xi_k-E_F|}.
\end{align}
Here let us assume that $\tau_q$ is longer than the peak time ($t^*$).
When $|\xi_k-E_F|=10$ GHz and $d=200$ nm, $\tau_q$ is of the order of
ns, which is consistent with the experimental result shown in 
Fig.~\ref{fig:exp_data}(b), where the electron peak time is of the order of ns,
at least.
If $d=1$ nm, $\tau_q$ shortens as ${\cal O}({\rm ps})$ and is inconsistent
with the experiment.
Since $\tau_q$ is independent of the charge,
a unique solution for explaining $v \ll v_F$ 
is to assume $e_*$ (instead of $e$) as shown in Eq.~(\ref{eq:new_vq})
and use Eq.~(\ref{eq:new_vq}) with Eq.~(\ref{app:1}).

We note that $t^*$ should not be identified with
the transport relaxation time ($\tau$), 
which is estimated from the mobility $\mu$
using $\mu=e_*\tau/m$ where the effective mass $m$ satisfies $m v_F^2/2=|E_F|$.
Because, when $E_F=0.1$ eV, $\mu\simeq 5000$ cm$^2$/V$\cdot$s is the
typical value for epitaxial graphene samples,~\cite{takase12} 
$\tau$ is the order of picoseconds.
This result is not in good agreement with the experiment showing
that $t^*$ is the order of nanoseconds.
Even though the Coulomb (electron-electron) interaction provides a
finite quasi-particle lifetime, it does not contribute to the transport
time.
We also note that the plasmon lifetime determined by the Coulomb
interaction is estimated in Refs.~\onlinecite{Principi2013a}
and~\onlinecite{Principi2013}.

\bibliographystyle{apsrev4-1}
%

\begin{thebibliography}{28}%
\makeatletter
\providecommand \@ifxundefined [1]{%
 \@ifx{#1\undefined}
}%
\providecommand \@ifnum [1]{%
 \ifnum #1\expandafter \@firstoftwo
 \else \expandafter \@secondoftwo
 \fi
}%
\providecommand \@ifx [1]{%
 \ifx #1\expandafter \@firstoftwo
 \else \expandafter \@secondoftwo
 \fi
}%
\providecommand \natexlab [1]{#1}%
\providecommand \enquote  [1]{``#1''}%
\providecommand \bibnamefont  [1]{#1}%
\providecommand \bibfnamefont [1]{#1}%
\providecommand \citenamefont [1]{#1}%
\providecommand \href@noop [0]{\@secondoftwo}%
\providecommand \href [0]{\begingroup \@sanitize@url \@href}%
\providecommand \@href[1]{\@@startlink{#1}\@@href}%
\providecommand \@@href[1]{\endgroup#1\@@endlink}%
\providecommand \@sanitize@url [0]{\catcode `\\12\catcode `\$12\catcode
  `\&12\catcode `\#12\catcode `\^12\catcode `\_12\catcode `\%12\relax}%
\providecommand \@@startlink[1]{}%
\providecommand \@@endlink[0]{}%
\providecommand \url  [0]{\begingroup\@sanitize@url \@url }%
\providecommand \@url [1]{\endgroup\@href {#1}{\urlprefix }}%
\providecommand \urlprefix  [0]{URL }%
\providecommand \Eprint [0]{\href }%
\@ifxundefined \urlstyle {%
  \providecommand \doi  [0]{\begingroup \@sanitize@url \@doi}%
  \providecommand \@doi [1]{\endgroup \@@startlink {\doibase
  #1}doi:\discretionary {}{}{}#1\@@endlink }%
}{%
  \providecommand \doi  [0]{doi:\discretionary{}{}{}\begingroup
  \urlstyle{rm}\Url }%
}%
\providecommand \doibase [0]{http://dx.doi.org/}%
\providecommand \Doi [0]{\begingroup \@sanitize@url \@Doi }%
\providecommand \@Doi  [1]{\endgroup\@@startlink{\doibase#1}\@@Doi}%
\providecommand \@@Doi [1]{#1\@@endlink}%
\providecommand \selectlanguage [0]{\@gobble}%
\providecommand \bibinfo  [0]{\@secondoftwo}%
\providecommand \bibfield  [0]{\@secondoftwo}%
\providecommand \translation [1]{[#1]}%
\providecommand \BibitemOpen [0]{}%
\providecommand \bibitemStop [0]{}%
\providecommand \bibitemNoStop [0]{.\EOS\space}%
\providecommand \EOS [0]{\spacefactor3000\relax}%
\providecommand \BibitemShut  [1]{\csname bibitem#1\endcsname}%
\bibitem [{\citenamefont {Mahan}(2000)}]{Mahan2000}%
  \BibitemOpen
  \bibfield  {author} {\bibinfo {author} {\bibfnamefont {G.~D.}\ \bibnamefont
  {Mahan}},\ }\href@noop {} {\emph {\bibinfo {title} {{Many-Particle
  Physics}}}}\ (\bibinfo  {publisher} {Springer},\ \bibinfo {year}
  {2000})\BibitemShut {NoStop}%
\bibitem [{\citenamefont {Ando}(1982)}]{Ando1982}%
  \BibitemOpen
  \bibfield  {author} {\bibinfo {author} {\bibfnamefont {T.}~\bibnamefont
  {Ando}},\ }\Doi {10.1103/RevModPhys.54.437} {\bibfield  {journal} {\bibinfo
  {journal} {Reviews of Modern Physics},\ }\textbf {\bibinfo {volume} {54}},\
  \bibinfo {pages} {437} (\bibinfo {year} {1982})},\ ISSN \bibinfo {issn}
  {0034-6861}\BibitemShut {NoStop}%
\bibitem [{\citenamefont {Novoselov}\ \emph {et~al.}(2005)\citenamefont
  {Novoselov}, \citenamefont {Geim}, \citenamefont {Morozov}, \citenamefont
  {Jiang}, \citenamefont {Katsnelson}, \citenamefont {Grigorieva},
  \citenamefont {Dubonos},\ and\ \citenamefont {Firsov}}]{novoselov05}%
  \BibitemOpen
  \bibfield  {author} {\bibinfo {author} {\bibfnamefont {K.~S.}\ \bibnamefont
  {Novoselov}}, \bibinfo {author} {\bibfnamefont {A.~K.}\ \bibnamefont {Geim}},
  \bibinfo {author} {\bibfnamefont {S.~V.}\ \bibnamefont {Morozov}}, \bibinfo
  {author} {\bibfnamefont {D.}~\bibnamefont {Jiang}}, \bibinfo {author}
  {\bibfnamefont {M.~I.}\ \bibnamefont {Katsnelson}}, \bibinfo {author}
  {\bibfnamefont {I.~V.}\ \bibnamefont {Grigorieva}}, \bibinfo {author}
  {\bibfnamefont {S.~V.}\ \bibnamefont {Dubonos}}, \ and\ \bibinfo {author}
  {\bibfnamefont {A.~A.}\ \bibnamefont {Firsov}},\ }\href@noop {} {\bibfield
  {journal} {\bibinfo  {journal} {Nature},\ }\textbf {\bibinfo {volume}
  {438}},\ \bibinfo {pages} {197} (\bibinfo {year} {2005})}\BibitemShut
  {NoStop}%
\bibitem [{\citenamefont {Zhang}\ \emph {et~al.}(2005)\citenamefont {Zhang},
  \citenamefont {Tan}, \citenamefont {Stormer},\ and\ \citenamefont
  {Kim}}]{zhang05}%
  \BibitemOpen
  \bibfield  {author} {\bibinfo {author} {\bibfnamefont {Y.}~\bibnamefont
  {Zhang}}, \bibinfo {author} {\bibfnamefont {Y.-W.}\ \bibnamefont {Tan}},
  \bibinfo {author} {\bibfnamefont {H.~L.}\ \bibnamefont {Stormer}}, \ and\
  \bibinfo {author} {\bibfnamefont {P.}~\bibnamefont {Kim}},\ }\href@noop {}
  {\bibfield  {journal} {\bibinfo  {journal} {Nature},\ }\textbf {\bibinfo
  {volume} {438}},\ \bibinfo {pages} {201} (\bibinfo {year}
  {2005})}\BibitemShut {NoStop}%
\bibitem [{\citenamefont {Kumada}\ \emph {et~al.}(2014)\citenamefont {Kumada},
  \citenamefont {Dubourget}, \citenamefont {Sasaki}, \citenamefont {Tanabe},
  \citenamefont {Hibino}, \citenamefont {Kamata}, \citenamefont {Hashisaka},
  \citenamefont {Muraki},\ and\ \citenamefont {Fujisawa}}]{Kumada2014}%
  \BibitemOpen
  \bibfield  {author} {\bibinfo {author} {\bibfnamefont {N.}~\bibnamefont
  {Kumada}}, \bibinfo {author} {\bibfnamefont {R.}~\bibnamefont {Dubourget}},
  \bibinfo {author} {\bibfnamefont {K.}~\bibnamefont {Sasaki}}, \bibinfo
  {author} {\bibfnamefont {S.}~\bibnamefont {Tanabe}}, \bibinfo {author}
  {\bibfnamefont {H.}~\bibnamefont {Hibino}}, \bibinfo {author} {\bibfnamefont
  {H.}~\bibnamefont {Kamata}}, \bibinfo {author} {\bibfnamefont
  {M.}~\bibnamefont {Hashisaka}}, \bibinfo {author} {\bibfnamefont
  {K.}~\bibnamefont {Muraki}}, \ and\ \bibinfo {author} {\bibfnamefont
  {T.}~\bibnamefont {Fujisawa}},\ }\Doi {10.1088/1367-2630/16/6/063055}
  {\bibfield  {journal} {\bibinfo  {journal} {New Journal of Physics},\
  }\textbf {\bibinfo {volume} {16}},\ \bibinfo {pages} {063055} (\bibinfo
  {year} {2014})},\ ISSN \bibinfo {issn} {1367-2630}\BibitemShut {NoStop}%
\bibitem [{\citenamefont {Kumada}\ \emph {et~al.}(2013)\citenamefont {Kumada},
  \citenamefont {Tanabe}, \citenamefont {Hibino}, \citenamefont {Kamata},
  \citenamefont {Hashisaka}, \citenamefont {Muraki},\ and\ \citenamefont
  {Fujisawa}}]{Kumada2013}%
  \BibitemOpen
  \bibfield  {author} {\bibinfo {author} {\bibfnamefont {N.}~\bibnamefont
  {Kumada}}, \bibinfo {author} {\bibfnamefont {S.}~\bibnamefont {Tanabe}},
  \bibinfo {author} {\bibfnamefont {H.}~\bibnamefont {Hibino}}, \bibinfo
  {author} {\bibfnamefont {H.}~\bibnamefont {Kamata}}, \bibinfo {author}
  {\bibfnamefont {M.}~\bibnamefont {Hashisaka}}, \bibinfo {author}
  {\bibfnamefont {K.}~\bibnamefont {Muraki}}, \ and\ \bibinfo {author}
  {\bibfnamefont {T.}~\bibnamefont {Fujisawa}},\ }\Doi {10.1038/ncomms2353}
  {\bibfield  {journal} {\bibinfo  {journal} {Nature communications},\ }\textbf
  {\bibinfo {volume} {4}},\ \bibinfo {pages} {1363} (\bibinfo {year} {2013})},\
  ISSN \bibinfo {issn} {2041-1723}\BibitemShut {NoStop}%
\bibitem [{\citenamefont {Klimchitskaya}\ \emph {et~al.}(2014)\citenamefont
  {Klimchitskaya}, \citenamefont {Mostepanenko},\ and\ \citenamefont
  {Sernelius}}]{Klimchitskaya2014}%
  \BibitemOpen
  \bibfield  {author} {\bibinfo {author} {\bibfnamefont {G.~L.}\ \bibnamefont
  {Klimchitskaya}}, \bibinfo {author} {\bibfnamefont {V.~M.}\ \bibnamefont
  {Mostepanenko}}, \ and\ \bibinfo {author} {\bibfnamefont {B.~E.}\
  \bibnamefont {Sernelius}},\ }\Doi {10.1103/PhysRevB.89.125407} {\bibfield
  {journal} {\bibinfo  {journal} {Physical Review B},\ }\textbf {\bibinfo
  {volume} {89}},\ \bibinfo {pages} {125407} (\bibinfo {year} {2014})},\ ISSN
  \bibinfo {issn} {1098-0121}\BibitemShut {NoStop}%
\bibitem [{\citenamefont {Wunsch}\ \emph {et~al.}(2006)\citenamefont {Wunsch},
  \citenamefont {Stauber}, \citenamefont {Sols},\ and\ \citenamefont
  {Guinea}}]{Wunsch2006}%
  \BibitemOpen
  \bibfield  {author} {\bibinfo {author} {\bibfnamefont {B.}~\bibnamefont
  {Wunsch}}, \bibinfo {author} {\bibfnamefont {T.}~\bibnamefont {Stauber}},
  \bibinfo {author} {\bibfnamefont {F.}~\bibnamefont {Sols}}, \ and\ \bibinfo
  {author} {\bibfnamefont {F.}~\bibnamefont {Guinea}},\ }\Doi
  {10.1088/1367-2630/8/12/318} {\bibfield  {journal} {\bibinfo  {journal} {New
  Journal of Physics},\ }\textbf {\bibinfo {volume} {8}},\ \bibinfo {pages}
  {318} (\bibinfo {year} {2006})},\ ISSN \bibinfo {issn}
  {1367-2630}\BibitemShut {NoStop}%
\bibitem [{\citenamefont {Hwang}\ and\ \citenamefont {{Das
  Sarma}}(2007)}]{hwang07}%
  \BibitemOpen
  \bibfield  {author} {\bibinfo {author} {\bibfnamefont {E.~H.}\ \bibnamefont
  {Hwang}}\ and\ \bibinfo {author} {\bibfnamefont {S.}~\bibnamefont {{Das
  Sarma}}},\ }\Doi {10.1103/PhysRevB.75.205418} {\bibfield  {journal} {\bibinfo
   {journal} {Physical Review B},\ }\textbf {\bibinfo {volume} {75}},\ \bibinfo
  {pages} {205418} (\bibinfo {year} {2007})},\ ISSN \bibinfo {issn}
  {1098-0121}\BibitemShut {NoStop}%
\bibitem [{\citenamefont {Sasaki}\ \emph {et~al.}(2012)\citenamefont {Sasaki},
  \citenamefont {Kato}, \citenamefont {Tokura}, \citenamefont {Suzuki},\ and\
  \citenamefont {Sogawa}}]{sasaki12_migration}%
  \BibitemOpen
  \bibfield  {author} {\bibinfo {author} {\bibfnamefont {K.-i.}\ \bibnamefont
  {Sasaki}}, \bibinfo {author} {\bibfnamefont {K.}~\bibnamefont {Kato}},
  \bibinfo {author} {\bibfnamefont {Y.}~\bibnamefont {Tokura}}, \bibinfo
  {author} {\bibfnamefont {S.}~\bibnamefont {Suzuki}}, \ and\ \bibinfo {author}
  {\bibfnamefont {T.}~\bibnamefont {Sogawa}},\ }\href@noop {} {\bibfield
  {journal} {\bibinfo  {journal} {Phys. Rev. B},\ }\textbf {\bibinfo {volume}
  {86}},\ \bibinfo {pages} {201403} (\bibinfo {year} {2012})}\BibitemShut
  {NoStop}%
\bibitem [{\citenamefont {Stern}(1967)}]{stern67}%
  \BibitemOpen
  \bibfield  {author} {\bibinfo {author} {\bibfnamefont {F.}~\bibnamefont
  {Stern}},\ }\Doi {10.1103/PhysRevLett.18.546} {\bibfield  {journal} {\bibinfo
   {journal} {Physical Review Letters},\ }\textbf {\bibinfo {volume} {18}},\
  \bibinfo {pages} {546} (\bibinfo {year} {1967})},\ ISSN \bibinfo {issn}
  {0031-9007}\BibitemShut {NoStop}%
\bibitem [{\citenamefont {Nakayama}(1974)}]{nakayama74}%
  \BibitemOpen
  \bibfield  {author} {\bibinfo {author} {\bibfnamefont {M.}~\bibnamefont
  {Nakayama}},\ }\Doi {10.1143/JPSJ.36.393} {\bibfield  {journal} {\bibinfo
  {journal} {J. Phys. Soc. Jpn.},\ }\textbf {\bibinfo {volume} {36}},\ \bibinfo
  {pages} {393} (\bibinfo {year} {1974})}\BibitemShut {NoStop}%
\bibitem [{\citenamefont {Ando}(2008)}]{Ando2008}%
  \BibitemOpen
  \bibfield  {author} {\bibinfo {author} {\bibfnamefont {T.}~\bibnamefont
  {Ando}},\ }\Doi {10.1143/PTPS.176.203} {\bibfield  {journal} {\bibinfo
  {journal} {Progress of Theoretical Physics Supplement},\ }\textbf {\bibinfo
  {volume} {176}},\ \bibinfo {pages} {203} (\bibinfo {year} {2008})},\ ISSN
  \bibinfo {issn} {0375-9687}\BibitemShut {NoStop}%
\bibitem [{\citenamefont {Zebrev}(2011)}]{Zebrev2011}%
  \BibitemOpen
  \bibfield  {author} {\bibinfo {author} {\bibfnamefont {G.}~\bibnamefont
  {Zebrev}},\ }in\ \Doi {10.5772/14211} {\emph {\bibinfo {booktitle} {Physics
  and Applications of Graphene - Theory}}},\ \bibinfo {editor} {edited by\
  \bibinfo {editor} {\bibfnamefont {S.}~\bibnamefont {Mikhailov}}}\ (\bibinfo
  {publisher} {InTech},\ \bibinfo {year} {2011})\ p.\ \bibinfo {pages}
  {475}\BibitemShut {NoStop}%
\bibitem [{\citenamefont {Takase}\ \emph {et~al.}(2012)\citenamefont {Takase},
  \citenamefont {Tanabe}, \citenamefont {Sasaki}, \citenamefont {Hibino},\ and\
  \citenamefont {Muraki}}]{takase12}%
  \BibitemOpen
  \bibfield  {author} {\bibinfo {author} {\bibfnamefont {K.}~\bibnamefont
  {Takase}}, \bibinfo {author} {\bibfnamefont {S.}~\bibnamefont {Tanabe}},
  \bibinfo {author} {\bibfnamefont {S.}~\bibnamefont {Sasaki}}, \bibinfo
  {author} {\bibfnamefont {H.}~\bibnamefont {Hibino}}, \ and\ \bibinfo {author}
  {\bibfnamefont {K.}~\bibnamefont {Muraki}},\ }\href@noop {} {\bibfield
  {journal} {\bibinfo  {journal} {Phys. Rev. B},\ }\textbf {\bibinfo {volume}
  {86}},\ \bibinfo {pages} {165435} (\bibinfo {year} {2012})}\BibitemShut
  {NoStop}%
\bibitem [{\citenamefont {Luryi}(1988)}]{Luryi1988}%
  \BibitemOpen
  \bibfield  {author} {\bibinfo {author} {\bibfnamefont {S.}~\bibnamefont
  {Luryi}},\ }\Doi {10.1063/1.99649} {\bibfield  {journal} {\bibinfo  {journal}
  {Applied Physics Letters},\ }\textbf {\bibinfo {volume} {52}},\ \bibinfo
  {pages} {501} (\bibinfo {year} {1988})},\ ISSN \bibinfo {issn}
  {00036951}\BibitemShut {NoStop}%
\bibitem [{\citenamefont {Burke}\ \emph {et~al.}(2000)\citenamefont {Burke},
  \citenamefont {Spielman}, \citenamefont {Eisenstein}, \citenamefont
  {Pfeiffer},\ and\ \citenamefont {West}}]{burke00}%
  \BibitemOpen
  \bibfield  {author} {\bibinfo {author} {\bibfnamefont {P.~J.}\ \bibnamefont
  {Burke}}, \bibinfo {author} {\bibfnamefont {I.~B.}\ \bibnamefont {Spielman}},
  \bibinfo {author} {\bibfnamefont {J.~P.}\ \bibnamefont {Eisenstein}},
  \bibinfo {author} {\bibfnamefont {L.~N.}\ \bibnamefont {Pfeiffer}}, \ and\
  \bibinfo {author} {\bibfnamefont {K.~W.}\ \bibnamefont {West}},\ }\Doi
  {10.1063/1.125881} {\bibfield  {journal} {\bibinfo  {journal} {Appl. Phys.
  Lett.},\ }\textbf {\bibinfo {volume} {76}},\ \bibinfo {pages} {745} (\bibinfo
  {year} {2000})}\BibitemShut {NoStop}%
\bibitem [{\citenamefont {Doetsch}(1971)}]{doetsch71}%
  \BibitemOpen
  \bibfield  {author} {\bibinfo {author} {\bibfnamefont {G.}~\bibnamefont
  {Doetsch}},\ }\href@noop {} {\emph {\bibinfo {title} {{Guide to the
  applications of the Laplace and Z-transforms}}}}\ (\bibinfo  {publisher}
  {Reinhold},\ \bibinfo {year} {1971})\BibitemShut {NoStop}%
\bibitem [{\citenamefont {Sonnenschein}\ \emph {et~al.}(2000)\citenamefont
  {Sonnenschein}, \citenamefont {Rutkevich},\ and\ \citenamefont
  {Censor}}]{sonnenschein00}%
  \BibitemOpen
  \bibfield  {author} {\bibinfo {author} {\bibfnamefont {E.}~\bibnamefont
  {Sonnenschein}}, \bibinfo {author} {\bibfnamefont {I.}~\bibnamefont
  {Rutkevich}}, \ and\ \bibinfo {author} {\bibfnamefont {D.}~\bibnamefont
  {Censor}},\ }\href@noop {} {\bibfield  {journal} {\bibinfo  {journal}
  {Progress In Electromagnetics Research},\ }\textbf {\bibinfo {volume} {27}},\
  \bibinfo {pages} {129} (\bibinfo {year} {2000})}\BibitemShut {NoStop}%
\bibitem [{\citenamefont {Quinn}\ and\ \citenamefont
  {Ferrell}(1958)}]{Quinn1958}%
  \BibitemOpen
  \bibfield  {author} {\bibinfo {author} {\bibfnamefont {J.}~\bibnamefont
  {Quinn}}\ and\ \bibinfo {author} {\bibfnamefont {R.}~\bibnamefont
  {Ferrell}},\ }\Doi {10.1103/PhysRev.112.812} {\bibfield  {journal} {\bibinfo
  {journal} {Physical Review},\ }\textbf {\bibinfo {volume} {112}},\ \bibinfo
  {pages} {812} (\bibinfo {year} {1958})},\ ISSN \bibinfo {issn}
  {0031-899X}\BibitemShut {NoStop}%
\bibitem [{\citenamefont {Hawrylak}(1987)}]{Hawrylak1987}%
  \BibitemOpen
  \bibfield  {author} {\bibinfo {author} {\bibfnamefont {P.}~\bibnamefont
  {Hawrylak}},\ }\Doi {10.1103/PhysRevLett.59.485} {\bibfield  {journal}
  {\bibinfo  {journal} {Physical Review Letters},\ }\textbf {\bibinfo {volume}
  {59}},\ \bibinfo {pages} {485} (\bibinfo {year} {1987})},\ ISSN \bibinfo
  {issn} {0031-9007}\BibitemShut {NoStop}%
\bibitem [{\citenamefont {{Das Sarma}}\ \emph {et~al.}(2007)\citenamefont {{Das
  Sarma}}, \citenamefont {Hwang},\ and\ \citenamefont {Tse}}]{DasSarma2007}%
  \BibitemOpen
  \bibfield  {author} {\bibinfo {author} {\bibfnamefont {S.}~\bibnamefont {{Das
  Sarma}}}, \bibinfo {author} {\bibfnamefont {E.}~\bibnamefont {Hwang}}, \ and\
  \bibinfo {author} {\bibfnamefont {W.-K.}\ \bibnamefont {Tse}},\ }\Doi
  {10.1103/PhysRevB.75.121406} {\bibfield  {journal} {\bibinfo  {journal}
  {Physical Review B},\ }\textbf {\bibinfo {volume} {75}},\ \bibinfo {pages}
  {121406} (\bibinfo {year} {2007})},\ ISSN \bibinfo {issn}
  {1098-0121}\BibitemShut {NoStop}%
\bibitem [{\citenamefont {Principi}\ \emph {et~al.}(2011)\citenamefont
  {Principi}, \citenamefont {Asgari},\ and\ \citenamefont
  {Polini}}]{Principi2011}%
  \BibitemOpen
  \bibfield  {author} {\bibinfo {author} {\bibfnamefont {A.}~\bibnamefont
  {Principi}}, \bibinfo {author} {\bibfnamefont {R.}~\bibnamefont {Asgari}}, \
  and\ \bibinfo {author} {\bibfnamefont {M.}~\bibnamefont {Polini}},\ }\Doi
  {10.1016/j.ssc.2011.07.015} {\bibfield  {journal} {\bibinfo  {journal} {Solid
  State Communications},\ }\textbf {\bibinfo {volume} {151}},\ \bibinfo {pages}
  {1627} (\bibinfo {year} {2011})},\ ISSN \bibinfo {issn}
  {00381098}\BibitemShut {NoStop}%
\bibitem [{\citenamefont {Principi}\ \emph
  {et~al.}(2013){\natexlab{a}}\citenamefont {Principi}, \citenamefont
  {Vignale}, \citenamefont {Carrega},\ and\ \citenamefont
  {Polini}}]{Principi2013a}%
  \BibitemOpen
  \bibfield  {author} {\bibinfo {author} {\bibfnamefont {A.}~\bibnamefont
  {Principi}}, \bibinfo {author} {\bibfnamefont {G.}~\bibnamefont {Vignale}},
  \bibinfo {author} {\bibfnamefont {M.}~\bibnamefont {Carrega}}, \ and\
  \bibinfo {author} {\bibfnamefont {M.}~\bibnamefont {Polini}},\ }\Doi
  {10.1103/PhysRevB.88.121405} {\bibfield  {journal} {\bibinfo  {journal}
  {Physical Review B},\ }\textbf {\bibinfo {volume} {88}},\ \bibinfo {pages}
  {121405} (\bibinfo {year} {2013}{\natexlab{a}})},\ ISSN \bibinfo {issn}
  {1098-0121}\BibitemShut {NoStop}%
\bibitem [{\citenamefont {Principi}\ \emph
  {et~al.}(2013){\natexlab{b}}\citenamefont {Principi}, \citenamefont
  {Vignale}, \citenamefont {Carrega},\ and\ \citenamefont
  {Polini}}]{Principi2013}%
  \BibitemOpen
  \bibfield  {author} {\bibinfo {author} {\bibfnamefont {A.}~\bibnamefont
  {Principi}}, \bibinfo {author} {\bibfnamefont {G.}~\bibnamefont {Vignale}},
  \bibinfo {author} {\bibfnamefont {M.}~\bibnamefont {Carrega}}, \ and\
  \bibinfo {author} {\bibfnamefont {M.}~\bibnamefont {Polini}},\ }\Doi
  {10.1103/PhysRevB.88.195405} {\bibfield  {journal} {\bibinfo  {journal}
  {Physical Review B},\ }\textbf {\bibinfo {volume} {88}},\ \bibinfo {pages}
  {195405} (\bibinfo {year} {2013}{\natexlab{b}})},\ ISSN \bibinfo {issn}
  {1098-0121}\BibitemShut {NoStop}%
\bibitem [{\citenamefont {Mikhailov}\ and\ \citenamefont
  {Ziegler}(2007)}]{Mikhailov2007}%
  \BibitemOpen
  \bibfield  {author} {\bibinfo {author} {\bibfnamefont {S.}~\bibnamefont
  {Mikhailov}}\ and\ \bibinfo {author} {\bibfnamefont {K.}~\bibnamefont
  {Ziegler}},\ }\Doi {10.1103/PhysRevLett.99.016803} {\bibfield  {journal}
  {\bibinfo  {journal} {Physical Review Letters},\ }\textbf {\bibinfo {volume}
  {99}},\ \bibinfo {pages} {016803} (\bibinfo {year} {2007})},\ ISSN \bibinfo
  {issn} {0031-9007}\BibitemShut {NoStop}%
\bibitem [{\citenamefont {Bordag}\ and\ \citenamefont
  {Pirozhenko}(2014)}]{Bordag2014}%
  \BibitemOpen
  \bibfield  {author} {\bibinfo {author} {\bibfnamefont {M.}~\bibnamefont
  {Bordag}}\ and\ \bibinfo {author} {\bibfnamefont {I.~G.}\ \bibnamefont
  {Pirozhenko}},\ }\Doi {10.1103/PhysRevB.89.035421} {\bibfield  {journal}
  {\bibinfo  {journal} {Physical Review B},\ }\textbf {\bibinfo {volume}
  {89}},\ \bibinfo {pages} {035421} (\bibinfo {year} {2014})},\ ISSN \bibinfo
  {issn} {1098-0121}\BibitemShut {NoStop}%
\bibitem [{\citenamefont {Allen}(1997)}]{Allen1997}%
  \BibitemOpen
  \bibfield  {author} {\bibinfo {author} {\bibfnamefont {P.~B.}\ \bibnamefont
  {Allen}},\ }\Doi {10.1007/BFb0106010} {\emph {\bibinfo {title} {{From Quantum
  Mechanics to Technology}}}},\ edited by\ \bibinfo {editor} {\bibfnamefont
  {Z.}~\bibnamefont {Petru}}, \bibinfo {editor} {\bibfnamefont
  {J.}~\bibnamefont {Przystawa}}, \ and\ \bibinfo {editor} {\bibfnamefont
  {K.}~\bibnamefont {Rapcewicz}},\ \bibinfo {series} {Lecture Notes in
  Physics}, Vol.\ \bibinfo {volume} {477}\ (\bibinfo  {publisher} {Springer
  Berlin Heidelberg},\ \bibinfo {year} {1997})\ ISBN \bibinfo {isbn}
  {978-3-540-61792-1},\ pp.\ \bibinfo {pages} {125--141}\BibitemShut {NoStop}%
\end{thebibliography}

%

\end{document}